\documentclass{aa}

\newcommand{\source}{MAXI~J1810$-$222}
\newcommand{\sou}{J1810}

\newcommand{\Msun}{$M_{\odot}$}

\newcommand{\swift}{\textsl{Swift}}
\newcommand{\xmm} {\textit{XMM-Newton}}
\newcommand{\nustar}{\textit{NuSTAR}}
\newcommand{\nicer}{\textit{NICER}}

\usepackage{graphicx}
\usepackage{color}
\usepackage{multirow}
\usepackage{txfonts}
\usepackage{hyperref}

\begin{document} 
   
   \title{A high-resolution X-ray view of the ultra-fast outflow \\ in MAXI\,J1810$-$222}

   \author{C. Pinto
          \inst{1}
          \and
          M. Del Santo
          \inst{1}
          \and 
          A. D'Aì
          \inst{1} 
          \and
          F. Pintore
          \inst{1}
          \and
          T. Russell
          \inst{1}
          \and          
          M. Parra
          \inst{2}
          \and
          J. Ferreira
          \inst{3}
          \and
          P.-O. Petrucci
          \inst{3}
          \and \\
          K. Fukumura
          \inst{4}
          \and 
          A. Marino
          \inst{5,6,1}
          \and 
          T. {Mu{\~n}oz-Darias}
          \inst{7,8}
          \and 
          G. A. Rodr{\'\i}guez Castillo
          \inst{1}
          \and 
          A. Segreto
          \inst{1}
          }

   \institute{INAF – IASF Palermo, Via U. La Malfa 153, I-90146 Palermo, Italy; 
              \email{ciro.pinto@inaf.it}
         \and
              Department of Physics, Ehime University, 2-5, Bunkyocho, Matsuyama, Ehime 790-8577, Japan
         \and
              Univ. Grenoble Alpes, CNRS, IPAG, 38000 Grenoble, France
         \and
              Department of Physics and Astronomy, James Madison University, Harrisonburg, VA
        \and
            Institute of Space Sciences (ICE), CSIC, Campus UAB, Barcelona, E-08193, Spain
        \and
            Institut d'Estudis Espacials de Catalunya (IEEC), 08860 Castelldefels (Barcelona), Spain 
        \and
            Instituto de Astrofísica de Canarias, E-38205 La Laguna, Tenerife, Spain
        \and
            Departamento de Astrofísica, Universidad de La Laguna, E-38206 La Laguna, Tenerife, Spain
             }

 \date{Received November 28, 2025; accepted March 16, 2026}
 
  \abstract
   {In previous work, it was reported that the Galactic black hole candidate MAXI J1810$-$222 exhibited a notable absorption spectral feature at around 1 keV in low-resolution X-ray spectra of CCD-like detectors. The feature was correlated with the spectral state of the source, being stronger in the soft states, as it occurs in the typical Fe\,K winds of X-ray binaries (XRBs). However, the results hinted towards rather extreme wind velocities of up to $\sim0.1\,c$.}
   {We therefore requested and obtained an observation with \textit{XMM-Newton} to take advantage of the {10-fold higher spectral resolution ($R = \lambda / \Delta \lambda \sim 200-400$)} provided by the RGS detector in order to resolve the lines and break the degeneracy between different models and interpretations.}
   {We applied state-of-the-art models of plasma in photoionisation equilibrium and multiphase interstellar medium. Further comparisons are performed with a re-analysis of \textit{NICER} and \textit{NuSTAR} data.}
   {The \textit{XMM-Newton}/RGS spectrum {is consistent with the presence of a mildly relativistic wind, confirming the earlier indications} obtained with \textit{NICER}, but places tighter constraints on the outflow properties, with the lines being intrinsically broad. {The data would then favour magnetically driven winds, although thermal effects may still contribute to mass loading.} \textit{NuSTAR} and \textit{XMM-Newton} (EPIC) show a further hotter component indicating a {stratified or multiphase} outflow. Fe\,K spectra taken with calorimetric detectors (e.g., Resolve on \textit{XRISM}) will enable a high-resolution view of the {complex extreme outflow in this source and shed new light on outflow processes in XRBs}.
}
   {}

   \keywords{accretion, accretion discs -- X-rays: binaries -- stars: winds, outflows -- X-rays: individual: MAXI J1810--222
               }

   \maketitle

\section{Introduction}

Black hole (BH) X-ray transients (BHTs) are binary systems composed of a BH accreting matter from a companion star. These objects spend most of their lifetimes in quiescence, where the X-ray luminosity is lower than $\sim$$10^{32}$ erg s$^{-1}$. However, when efficient accretion turns on, an outburst occurs, increasing the observed X-ray luminosity to $10^{36-39}$ erg s$^{-1}$. 
The outburst phase typically last from a few weeks up to several months \citep[see, e.g.,][]{tetarenko16}, although long outbursts lasting several years or even decades have been reported both in BH systems, such as GRS 1915+105 \citep[see, e.g.,][]{motta21, Deegan09}, GRO J1655$-$40 \citep{Sobczak99}, SWIFT J1753.5$-$0127 \citep{ZhangG19}, and in Low Mass X-ray Binaries (LMXBs) with NS, such as XMMU J174716.1$-$281048 \citep{Delsanto07} and 4U1608$-$52 \citep{Simon20}.

During a typical outburst, the X-ray spectra of BHTs usually show different states, such as the disc-dominated soft state and the Comptonisation dominated hard state, as well as intermediate states with spectral parameters in between these two canonical states (see, e.g. \citealt{Done2007} and references therein).
It is generally accepted that the soft and hard X-ray spectral components describe two different emitting regions, with the accretion disc dominating in the soft X-ray range and the hot accretion flow, i.e. the \textit{corona}, in the hard X-ray band (these components have comparable fluxes at around a few keV). 
However, these two regions are sufficiently close to interact with each other, giving rise to the reflection component due to the hot photons emitted by the corona being reprocessed by the accretion disc (see, e.g. \citealt{Garcia2020} and references therein).

Equatorial outflows in the form of disc winds are typically observed in disc-dominated states of sources viewed at high inclinations. This may depend on the radial density profile of the wind (the closer the line of sight is to the disc plane, the higher the optical depth, as discussed in \citealt{ponti12, Parra2024}).  
X-ray winds are mainly probed through resonant transitions of highly ionised elements that are blue-shifted by Doppler motions of a few hundred km/s. This indicates a lower limit on the wind escape radius on the order of 10$^4$--$10^{5}$ $R_g$. Here, the thermal motion of ions, due to the irradiation of X-ray photons from the inner disc, is sufficient to unbind them from the gravitational well of the compact object (\textit{thermally driven winds}, see e.g., \citealt{Tomaru2019} and references therein). 

{During hard states, winds with similar kinematical properties are detected in the optical and near-infrared bands (see, e.g., \citealt{MunozDarias2019}), suggesting that they are persistent and, at least in the most extreme cases, play a role in regulating the accretion flow (see, e.g., \citealt{MunozDarias2016}). However, they are only detectable in the X-ray band under favourable circumstances with respect to the outflow geometry or ionisation balance, suggesting possible stratification or multiphaseness (see, e.g., \citealt{MunozDarias2022}).}

In some BH systems, particularly those with higher wind speeds {being claimed}, \textit{magnetically driven winds} have been suggested as a better description due to their small launching radius  {\citep{Miller2006Nature,Chakravorty2016,Fukumura2017,Datta2024}}. The presence of a relativistic outflow in a sub-Eddington ($\dot{M} < 1\dot{M}_{\rm Edd}$) accreting stellar-mass black hole is considered strong evidence for a magnetically driven process because thermal gradients cannot drive winds faster than a few thousand km/s (see, e.g., \citealt{Done2018}). 

{Some X-ray binaries have shown evidence of powerful, relativistic or ultra-fast outflows (UFOs) blowing up to $0.1-0.2\,c$ but this is commonly observed at luminosities exceeding the Eddington limit as it occurs in photospheric bursts of accreting neutron stars (see, e.g., \citealt{Pinto2014b,Barra2025}) and ultra-luminous X-ray sources (ULXs, see, e.g. \citealt{Pinto2023a} and references therein). However, these cases are characterised by extremely soft X-ray spectra that do not fit within the typical spectral states framework of BHT systems {and are likely to enable different launching wind mechanisms}. Finally, it is worth noticing that UFOs are seen in a significant fraction of AGN, even at sub-Eddington rates ($\sim30$\,\%, see e.g. \citealt{Gianolli2024} and references therein). They often exhibit a cooler component in the soft X-ray band below 2\,keV, and show correlations between their properties and the underlying X-ray emitting continuum (see, e.g., \citealt{Xu2024}).}
  
\source\ (hereafter \sou) is an X-ray transient discovered in 2018 by MAXI/GSC  close to the Galactic Plane \citep{negoro18}. X-ray and radio spectral timing properties provided by follow-up observations with \swift/XRT, \swift/BAT and ATCA suggested that the source is most likely a new BHT {(see also Appendix\,\ref{sec:pulsations_search} for lack of pulsations)}, possibly located at a large distance {\citep[$\gtrsim8$\,kpc, see][hereafter R22]{russell22}.} 

Unlike most BHTs, \sou\ was discovered in a soft state and does not seem to follow the canonical hardness-intensity diagram (HID); instead, it goes back and forth from the hard state to the soft state several times (through intermediate states), making this source very peculiar.
Surprisingly, since its discovery 7 years ago, J1810 has remained active (e.g., \citealt{Marino2023a}), which makes it a member of the group of quasi-persistent X-ray binaries.
In a previous work \citep[][hereafter DS23]{MDS2023}, we exploited the \textit{NICER} observations of \sou\ performed in 2020 and found statistically significant spectral features around 1 keV in almost all spectra.
We confirmed that it was not instrumental by finding it also in a stacked spectrum with \swift/XRT and showing that it strongly depends on the spectral state of the source (DS23). The application of models of plasma in photoionisation equilibrium (PIE) allowed us to show that the plasma was outflowing with velocities up to $\sim0.1c$, with the speed peaking during the soft states.

Due to the limited spectral resolution of CCD-like spectra, we requested and were awarded a 60\,ks observation with \textit{XMM-Newton} in order to resolve and identify the spectral features through the reflection grating spectrometer (RGS).

\section{Observations and data reduction}

{In Table\,\ref{table:observations}, we briefly report the main observations used in this work along with the net exposure times. We avoid detail on the \textit{Neil Gehrels Swift Observatory} (hereafter \swift) as the full archival data set has been used.}

\begin{table}
\caption{Observations of J1810.}  
\label{table:observations}     
\renewcommand{\arraystretch}{1.5}
 \small\addtolength{\tabcolsep}{-4pt}
 \vspace{-0.1cm}
\scalebox{0.95}{
\hskip-0.0cm\begin{tabular}{@{}ccccccc}     
\hline  
Facility  & Obs\_ID  &  Date & \multicolumn{3}{c}{ $t_{\rm exp}$ (ks, instrument)} \\
  
\hline                                                                                                 
  \xmm\  & 0921250101 & 2023-09-16 & 57 (RGS) & 55 (pn) & 60 (MOS2) \\

  \multirow{2}{4em}{\textit{NICER}} & 6200560101 & 2023-09-15 & \multicolumn{3}{c}{\multirow{2}{4em}{3.4}} \\
                          & 6200560201 & 2023-09-16 &  \\   
\multirow{2}{4em}{NUSTAR} & 90410345001 & 2018-12-09 & \multicolumn{3}{c}{\multirow{2}{4em}{19}} \\
                          & 90402367002 & 2018-12-10 &  \\

\hline                          
\end{tabular}}

\vspace{0.3cm}

Notes: Exposure times account for the removal of periods of high background rate and passage for the South Atlantic Anomaly.
\end{table}

\subsection{Swift data}

\begin{figure}
   \centering
   \includegraphics[scale=0.32]{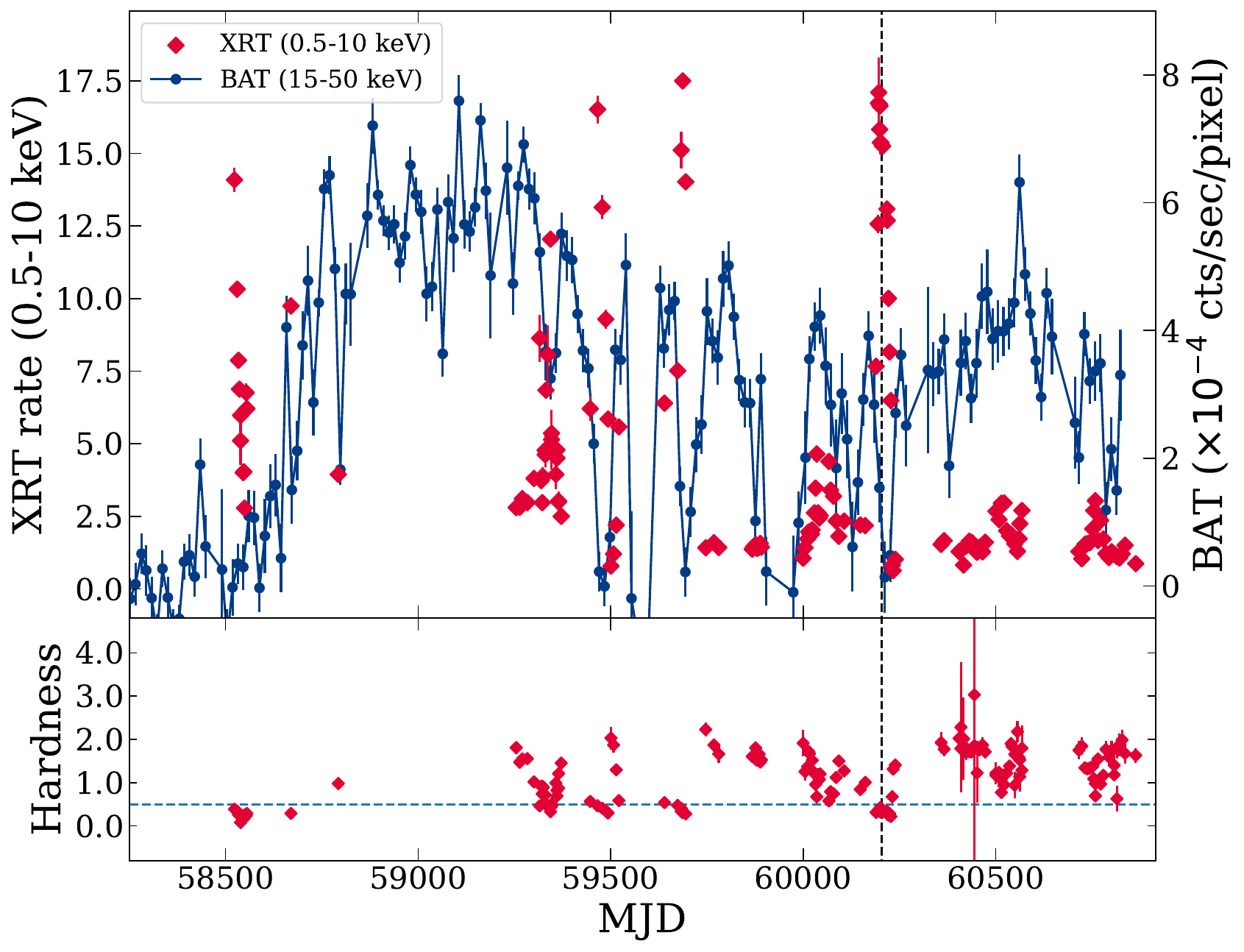}
\vspace{-0.3cm}
      \caption{Top panel: \swift/XRT (blue circles) and \swift/BAT (red diamonds) light curves. Bottom panel: XRT hardness defined as the ratio of the counts in the 2--10 keV and 0.5--2 keV energy bands. The dashed line indicates the \xmm\ observing time.}
         \label{fig:swift}
   \end{figure}

The outburst of MAXI J1810-222 has been regularly monitored by \swift-XRT, according to the telescope visibility. The observations shown in Fig.\ref{fig:swift} correspond to the full XRT dataset (target IDs 00011105 and 00016178), obtained between February 2019 and July 2025.
All observations were processed using the online Swift-XRT data products generator\footnote{\href{https://www.swift.ac.uk/user\_objects/}{https://www.swift.ac.uk/user\_objects/}}, developed by the UK Swift Science Data Centre. This generator relies on the latest software package and calibration files.
The light curve was extracted in the 0.5–10 keV energy range, while the hardness ratio was computed using the count rate in the 0.5–2 keV and 2–10 keV bands.

We used the \swift-XRT monitoring to trigger the XMM-Newton observation, setting the trigger condition to an observed XRT count rate above $\sim4$ ct/s and a hardness ratio (2-10 keV / 0.5-2 keV) lower than 0.4 (corresponding to high/intermediate and high/soft states). The XMM-Newton observation took place within a week from the trigger time. During the XMM-Newton observation, the XRT 0.5-10 keV count rate was approximately 17 ct/s, which corresponded to either the H1 or H2 (`high 1/2') spectra reported in DS23, where fast outflows were observed. During these two high states (soft or soft/intermediate), the temperature of the electron plasma was found to be around 20 keV when using hard X-ray data (see {R22}).

Data from the \swift-BAT survey were retrieved from the HEASARC public archive and processed using the BAT-IMAGER software \citep{segreto10}. This software is specifically designed for the analysis of data from coded-mask telescopes, generating all-sky maps by modelling and subtracting the background. Afterward, light curves and spectra can be extracted for any detected source.
For the case of MAXI J1810-222, we derived light curves in the energy range of 15-50 keV, with a time binning of 15 days (Fig.\ref{fig:swift}). During the \xmm\ observation the BAT count rate was very low, i.e. below $1 \times 10^{-4}$ ct/s/pixel, consistent with the source being in a soft state.

\subsection{\textit{XMM-Newton}} \label{sec:xmm}
\xmm\ observed \source\ on 16 September 2023 for a total exposure time of roughly 61 ks (ObsID 0921250101; P.I.: M. Del Santo).
All EPIC instruments were operated in TIMING mode with a thick filter. 
However, we discarded data from MOS1 due to an observation failure.
We processed raw data using the \textit{Science Analysis System} ({\scriptsize{SAS}}) version 21.0.0\,\footnote{\href{https://www.cosmos.esa.int/web/xmm-newton}{https://www.cosmos.esa.int/web/xmm-newton}} with most recent calibration files (as of April 2024) and \textsc{HeaSOFT} v. 6.34.  
{\scriptsize{EPPROC}} and {\scriptsize{EMPROC}} tasks have been used 
according to standard procedures regarding the patterns (FLAG==0 and PATTERN$\,\leq\,$4\,(12) for pn and MOS, respectively) and we checked for high particle background periods, that we spotted at the end of the observation. 
Following the standard threads for EPIC timing mode analysis\footnote{\href{https://www.cosmos.esa.int/web/xmm-newton/sas-threads}{https://www.cosmos.esa.int/web/xmm-newton/sas-threads}},
we selected source events from the central 29-47 (278-338) 
RAWX columns and background events from 3-5 (260-270) columns for pn (MOS2). For diagnostic checks, we also extracted an alternative {MOS2} background spectrum from the outer CCDs that collected data in IMAGING mode. We verified with the SAS tool {\sc{epatplot}} that the spectra are not affected by pile-up.
We extracted the light curves in two energy bands (0.5--2 and 2--10 keV) and calculated a hardness-ratio. No significant flux or spectral variability was observed.
Therefore, we extracted an averaged spectrum for the whole observation.
The response and ancillary files were extracted with the {\sc{rmfgen}} and {\sc{arfgen}} tasks.

A preliminary analysis of the pn and MOS2 spectra with simple models (disc black body and power-law) revealed instrumental artefacts, possibly due to incorrect background subtraction, incorrect gain or effective area calibration {(see Appendix\,\ref{sec:epic_calib})}.
After consulting with the \xmm\ calibration team (F. Fuerst, priv. comm.), 
we decided to turn off the default Rate-Dependent PHA (RDPHA) correction (withrdpha=’N’) and apply instead the Rate-Dependent CTI (RDCTI) correction, using {\sc{epfast}} (runepfast=’Y’).
However, the two methods did not produce any significant difference in the pn spectra, therefore we kept the standard processed data (i.e. with RDPHA). 

{In Appendix\,\ref{sec:pulsations_search}, we also show a search for periodic signals in the high-quality \xmm\ EPIC-pn data to confirm the non detection of coherent pulsations in this source in agreement with previous results (see, e.g., R22, DS23).}

The RGS data reduction was performed with the {\sc{rgsproc}} task which also extracts spectra and response/area files. We extracted the $1^{\rm st}$-order RGS spectra in a cross-dispersion region of 0.8' width, centred on the source coordinates and the background spectra by selecting photons
beyond the 98\% of the source point-spread function. We filtered out periods contaminated by high-background by selecting background-quiescent intervals in the light curves of the RGS 1,2 CCD\,9
({i.e.}, $\gtrsim1.7$\,keV) with a standard count rate below 0.2 cts s$^{-1}$. The RGS had a net exposure time of 57\,ks. Given the consistency between the spectra of the RGS 1 and 2 detectors, we produced a combined RGS spectrum through the {\sc{rgscombine}} task.

\subsection{\textit{NICER} data}

The Neutron Star Interior Composition Explorer (\textit{NICER}, \citealt{Gendreau2016}) observed the source during two separate days (Obsids 6200560101/0102) on 2023-09-15 and 2023-09-16.
We performed data reduction using {\sc{nicerdas}} software\footnote{\href{https://heasarc.gsfc.nasa.gov/docs/nicer/nicer_analysis.html}{https://heasarc.gsfc.nasa.gov/docs/nicer/nicer\_analysis.html}} version 11, along with \textit{NICER} {\sc{caldb}} xti20240206. 

For each observation, the vast majority of the exposure occurred during orbit `day' periods, which must be analysed independently and include significantly higher background contributions. A proper estimation of that background is only possible with the \texttt{scorpeon} background model, and this model requires knowledge of the geomagnetic data during the observations, which we  obtained using the {\sc{nicerdas}} task \texttt{nigeodown}.

We first reprocessed the observations using the {\sc{nicerdas}} task \texttt{nicerl2}, separating the orbit `day' and `night' periods and the different orbits, as high differences in calibration and background level require independent analysis. We then filtered for non X-ray flares using a combination of topological and variance based peak detection algorithms. Most notably, instead of the standard static overshoot rate (corresponding to all counts above 20\,keV, where the instrument itself has no effective area) threshold of 30 cts/s/FPM, we adopted a dynamical SNR base criterion and only excluded GTI periods with an overshoot rate $\gtrsim5$ times higher than the $2-8$\,keV count rate (computed on timescales of 1s). 
In a second step, we computed the spectral products of each gti period, in the 0.3-10\,keV range using \texttt{nicerl3-spect}. 
To ensure that no significant flare remained in the filtered GTIs, and to assess for any intrinsic variability of the source during the observations, we computed (and visually inspected) lightcurve products obtained using \texttt{nicerl3-lc}, with a 1s binning. The `day' data yields an exposure time of 3.4\,ks. 

In this work, we avoided performing a new reduction of the archival \nicer\ and \nustar\ data (which are compared with the new \nicer\ and \xmm\ data, see Table\,\ref{table:observations}) but simply retrieved them from previous work {(R22, DS23)} to obtain a consistent check of all results.

\section{Spectral modelling}

For spectral analysis, we used the {\scriptsize{SPEX}} fitting package 3.08.01 \citep{kaastraspex}, which has a full suite of models of line-emitting or absorbing plasmas. The spectra were grouped according to optimal binning \citep{Kaastra2016} directly in {\scriptsize{SPEX}} and fit by minimising the $C-$statistics \citep{Cash1979,Kaastra2017}.
This binning avoids oversampling the spectra by no more than a factor of three with respect to the spectral resolution and requires at least 1 count per noted bin to satisfy the requirements for the use of the Cash statistics. {All uncertainties are reported at the 68\% level, which is also the default in {\scriptsize{SPEX}}.}

\subsection{Baseline model} \label{sec:baseline}

Following DS23, we first focussed on obtaining a reasonable description of the broadband spectral continuum. At first, we modelled the \textit{XMM-Newton} and \textit{NICER} (day) spectrum taken between 0.5 keV and 10 keV. The spectral model is fitted simultaneously to the EPIC MOS2 / pn, RGS and \textit{NICER} spectra with a free multiplicative constant that accounts for the calibration uncertainties.  {The obvious and well-known mismatch between the EPIC CCD spectra in the timing mode (see also \citealt{Read2014}), as well as the instrumental artefacts mentioned in Sect.\,\ref{sec:xmm}, forced us to ignore their data below 2.6 keV (RGS and \textit{NICER} spectral shapes were instead compatible, see Fig.\,\ref{fig:epic_calib}). However, it is important to notice that both the EPIC MOS\,2 and pn data still confirm the presence of a strong, broad absorption feature at 1 keV (see Appendix\,\ref{sec:epic_calib})}. 

In analogy with DS23, we adopted a continuum model consisting of a disc-blackbody component (\texttt{dbb}) to account for the soft X-ray emission (peaking below 2 keV) and a Comptonised component (\texttt{comt}) describing the hard X-ray tail. We modelled the photo-electric absorption from the circumstellar and interstellar medium with the \texttt{hot} component, setting the temperature to $10^{-6}$ keV to describe a cold neutral gas. All abundances are reported in units of the recently  recommended Solar abundances of \citet{Lodders2009}, which are default in {\scriptsize{SPEX}}. In the {\scriptsize{SPEX}} formalism  the model reads as follows: \texttt{hot $\times$ (dbb + comt)}.

The Galactic coordinates of the source ($l=8.77, b=-1.98$) and a likely distance of 8 kpc, or larger, imply that our line of sight crosses the inner regions of the Milky Way, where the ISM abundances are subject to a significant gradient \citep{Pinto2013}. We therefore untied the metallicity of the \texttt{hot} component by coupling all elemental abundances to neon (for which we do not expect any depletion from the gas to the dust phase) and keeping the latter as a free parameter in the fit. The seed photon temperature of the \texttt{comt} was coupled to the \texttt{dbb} temperature, while its electron temperature was fixed to 20 keV following the \swift/BAT results for this intermediate-soft state (DS23).
Finally, we added an absorbing Gaussian line at 2.2 keV to account for an edge-like feature, most likely due to systematics in the \textit{NICER} effective area calibration (see, e.g., \citealt{Marino2023b}).

\begin{figure}
   \centering
   \includegraphics[scale=0.35]{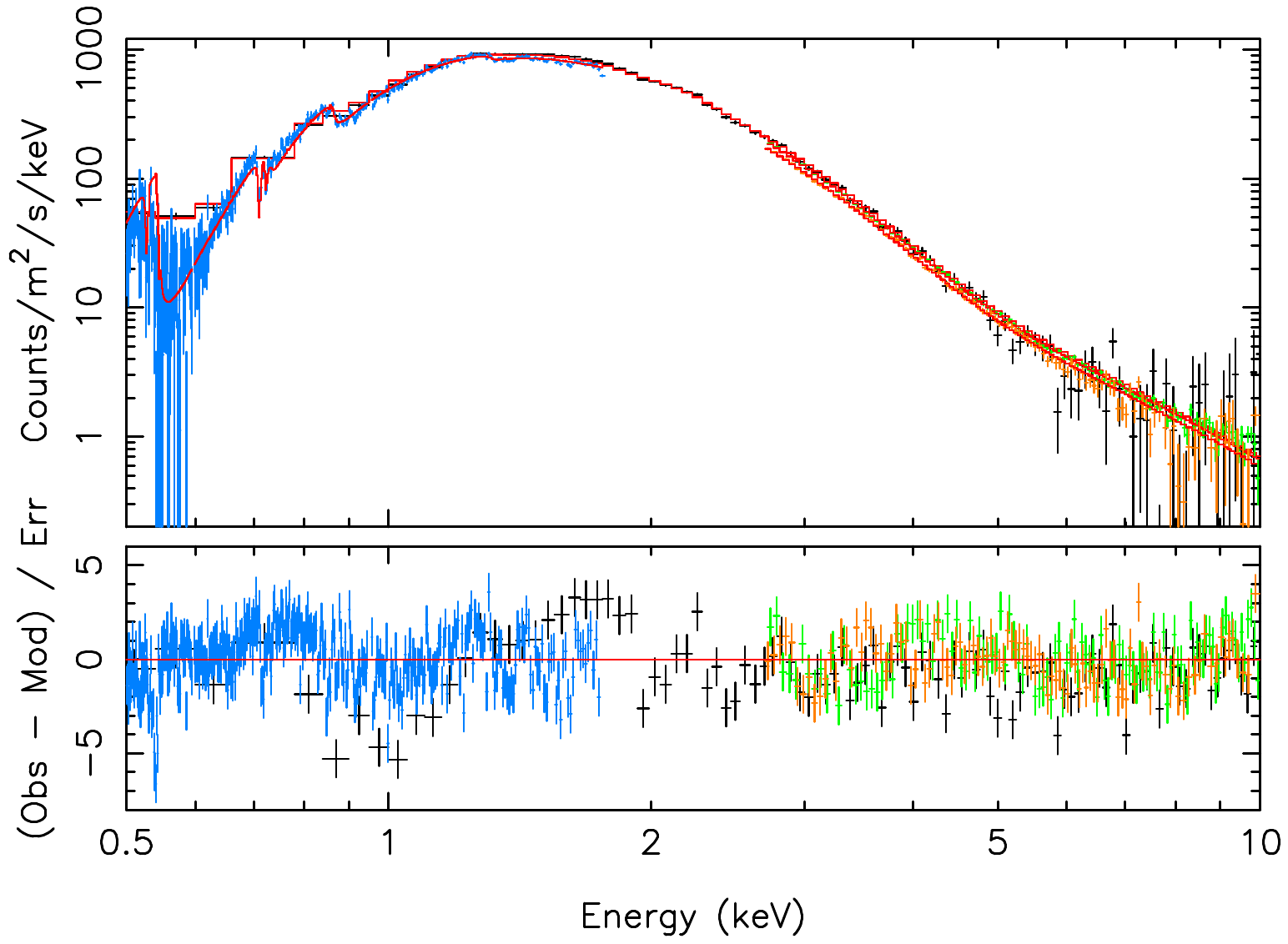}
    \vspace{-0.3cm}
      \caption{XMM / RGS (blue data) + EPIC (pn/green and MOS2/orange data) + \textit{NICER} (black data) spectra and  best-fit continuum model. The EPIC data were ignored below 2.6 keV due to calibration issues.}
         \label{fig:continuum_fit}
    \vspace{-0.2cm}
   \end{figure}

This simple continuum fit is shown in Fig.\,\ref{fig:continuum_fit}. The disc \texttt{dbb} temperature is around 0.9 keV, while the optical depth of the \texttt{comt} is about 0.4. Assuming a distance of 8 kpc, we measured an intrinsic 0.5-10\,keV total luminosity of $(9.1 \pm 0.3)\times10^{36}$\,erg\,s$^{-1}$ (for an observed or absorbed flux of $(3.9 \pm 0.1)\times10^{-10}$\,erg\,s$^{-1}$\,cm$^{-2}$) of which $\gtrsim90$\% is provided by the disc component.
The (neutral) interstellar medium has column density  $N_{\rm H , \, ISM}=(7.3 \pm 0.1)\times10^{21}$\,cm$^{-2}$ with a rather high metallicity of about {twice the Solar value}. These results are consistent with DS23. The \textit{NICER} spectrum shows a strong, broad absorption feature around 1 keV, which is confirmed by the shape of the RGS residuals. The fit statistics for this model are rather poor (C-stat of 1763 for 668 degrees of freedom, hereafter DOF) because we did not yet account for the ionised and dusty phases of the ISM and any XRB winds.

\subsection{Full ISM model}

In order to employ the high spectral resolution of the RGS, which is needed to accurately model the various interstellar phases and any leftovers due to the XRB winds, we ignored the \textit{NICER} data in the RGS operating band, i.e. below 1.77 keV. This is a common procedure for different types of source (see, e.g. \citealt{Pinto2021} and references therein). The \textit{NICER} spectrum yields a number of counts about 0.8 times with respect to the RGS spectrum {and, importantly, a much lower spectral resolution}, which would otherwise produce degeneracies between the different models, with poor constraints on line broadening. 

\begin{figure}
   \centering
   \includegraphics[scale=0.3]{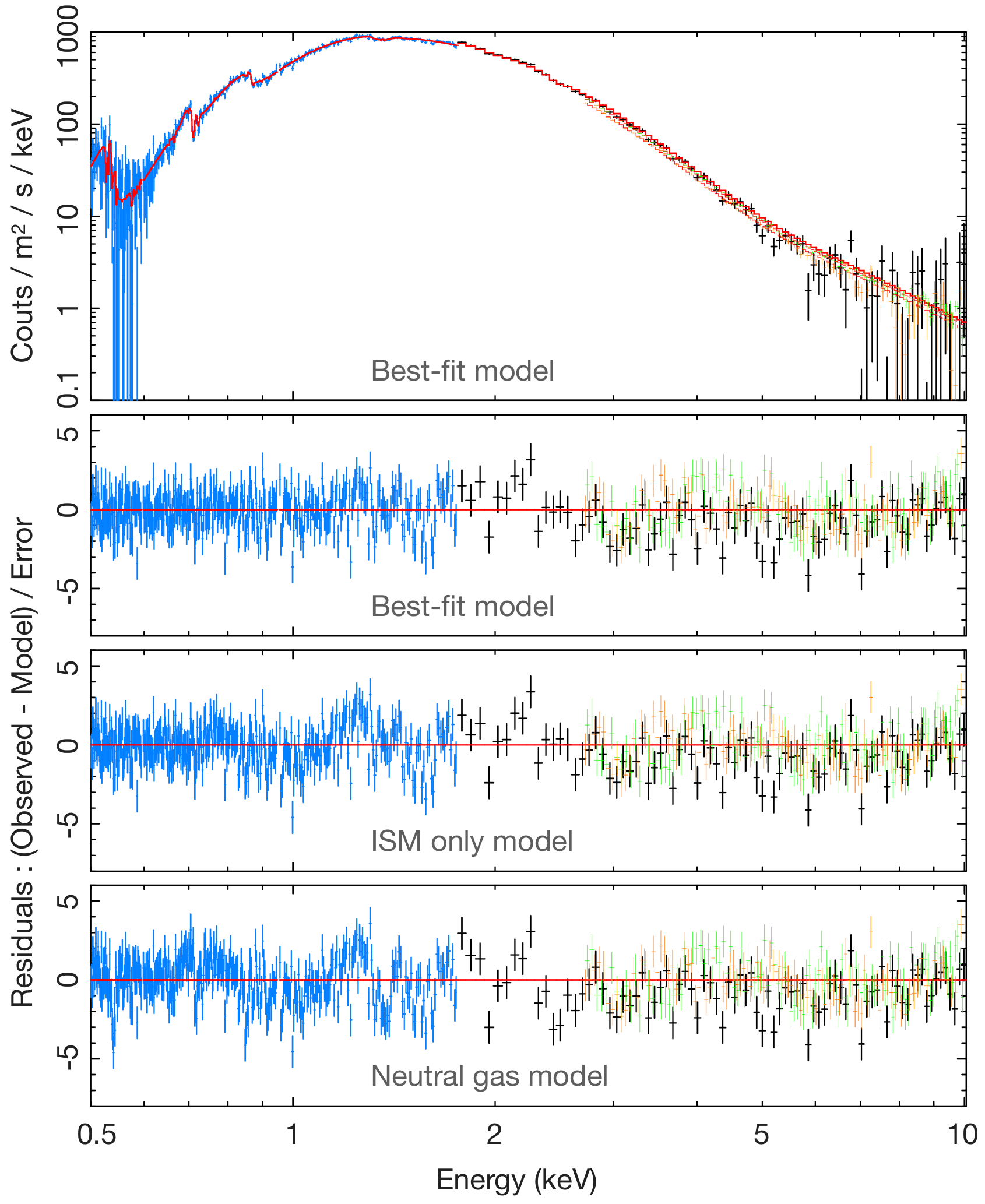}
      \caption{XMM / RGS + EPIC + \textit{NICER} spectra and best fit model (top two panels). The \textit{NICER} (EPIC) data were ignored between 0.5-1.8(2.6) keV in order to fully employ the high resolving power of RGS and decrease degeneracy between models of line emission / absorption as well as calibration issues. Bottom two panels:  residuals computed for the models without \texttt{pion} and without \texttt{pion}, ISM ions and dust.}
         \label{fig:best_fit}
   \end{figure}

A preliminary fit to the spectra with such a low-energy cut for \textit{NICER} (keeping all abundances coupled to neon, which is free to vary) provided a continuum model with parameters consistent with the previous fit {with a C-stat of 1454 for 648 DOF}. 
Since the RGS spectrum {resolves the individual interstellar K / L absorption edges, we left the abundances of oxygen, neon, magnesium, silicon and iron of the \texttt{hot} component free to vary} (with all other elements still coupled to neon).
The residuals for this model are shown in Fig.\,\ref{fig:best_fit} (bottom panel, dubbed `Neutral gas model'). The fit statistics are still poor {(C-stat of 1379 for 645 DOF, i.e., $\Delta$C-stat = 75 with in addition O, Mg and Fe abundances free to vary)} for the same reasons as mentioned above.

Through visual inspection of the narrow spectral residuals, we identified the $1s-2p$ transition energies of O\,{\sc ii-iv}, O\,{\sc vii-viii}, Ne\,{\sc ii-iii} and Ne\,{\sc ix-x} that are commonly observed in the ISM absorption spectra of X-ray background sources (see, e.g. \citealt{Gatuzz2018} and references therein). We found additional fine structures near the K-edges of oxygen (0.54\,keV), neon (0.87\,keV) and magnesium (1.31\,keV) and the L-edge of iron (0.71\,keV). The latter could be due to other weaker ionic lines (Ne\,{\sc v-vii}), dust (e.g. silicates, \citealt{Psaradaki2024,Rogantini2020}), and the photoionised wind previously claimed in this source or in other X-ray binaries (e.g. \citealt{Pinto2014b}).

Following \citet{Pinto2013}, we multiplied the continuum model for the \texttt{slab} model in {\scriptsize{SPEX}} which provides a complete set of ionic species for all elements with atomic number from 1 to 30. For both the \texttt{hot} and \texttt{slab} components, we adopted a reliable velocity dispersion of 100 km s$^{-1}$ {(ISM lines are too narrow anyway to be resolved with these data)}. Free in the fit were the column densities in the \texttt{slab} for the ionic species mentioned above. {This enables the ISM to account for any possible narrow rest-frame spectral feature present in the spectrum because the ionic column densities are not model dependent.}
The additional \texttt{slab} component significantly improves the spectral fit ($\Delta$C-stat = {97} for the inclusion of the 11 ionic species).

Finally, we added the contribution from solids, particularly for the modelling of the O, Fe and Mg edges, using the \texttt{amol} component, which contains a full suite of interstellar equivalents for terrestrial compounds that have been measured in the lab. We used a reliable composition of metallic iron and magnesium silicates (pyroxene, MgSiO$_3$), which so far has provided excellent descriptions of the K- and L-edges in the soft X-ray band \citep{Pinto2013,Rogantini2020,Psaradaki2024}. The column densities of these species were free parameters. This provided a major improvement of the edges and overall fit ($\Delta$C-stat / DOF = {37 / 2}). 

The spectral residuals for the full ISM model are shown in Fig.\,\ref{fig:best_fit} (dubbed `ISM only model'). Most narrow features have been taken into account, but there is still a broad feature around 1 keV. The fit statistics, although better, are still rather poor (C-stat / DOF = 1245 / 632).

\subsection{Photoionised plasma} \label{sec:bestfit}

Lastly, we included in the spectral model the photoionised plasma component. This was done using the \texttt{pion} model in {\scriptsize{SPEX}}; \texttt{pion} calculates the transmission and emission of a slab of photoionised plasma, where all ionic column densities are linked through a photoionisation model. The relevant parameter is the ionisation parameter $\xi = L/n_{\rm H}r^2$, with L the source luminosity, $n_{\rm H}$ the hydrogen density and $r$ the distance from the ionising source. At the moment, this is the only publicly available code that self-consistently computes the balance instantaneously during the spectral fitting. The ionising field or spectral energy distribution (SED) is, therefore the continuum model at each iteration. The final model can be described as follows:
$$\texttt{hot $\times$ slab $\times$ amol $\times$ pion $\times$ (dbb + comt)}.$$
The \texttt{pion} model can be used to produce spectra of PIE plasma both in emission and absorption. We adopted a pure-absorption model by setting a solid angle $\Omega/4\pi = 0$ and a covering fraction $f_{\rm cov} = 1$.
As additional free parameters, we chose the column density $N_{\rm H}$, the ionisation parameter $\xi$, the line-of-sight velocity $v_{\rm LOS}$, and the velocity dispersion or line width, $v_{\sigma}$. Given the limited statistics, we assumed Solar chemical abundances. This is generally a good approximation of low-counts spectra, although might have an impact on the detection of the winds given that their abundances may differ from the Solar pattern {(see, e.g., \citealt{Barra2024,Keshet2024,Kosec2025}).}

\begin{figure}
   \centering
   \includegraphics[scale=0.5]{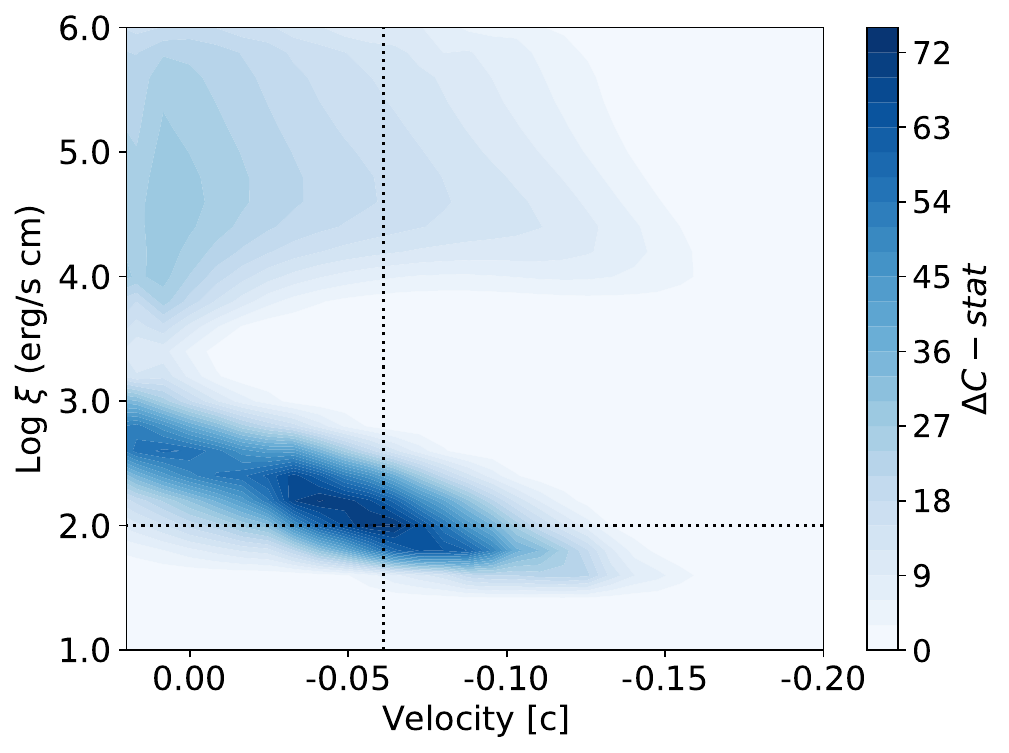}
    \vspace{-0.1cm}
      \caption{{Parameter-space scan throughout model grids of a photoionised plasma in absorption applied onto the \textit{XMM-Newton+NICER} spectra adopting a velocity dispersion of 20,000\,km/s. The dotted black lines identify the best-fit grid model. Structures are also present at high $\xi$.}}
         \label{fig:xmm_nicer_pion_scan}
    \vspace{-0.1cm}
   \end{figure}

Since we expected Doppler shifts and wanted to avoid getting stuck within a local minimum, we first performed a scan through model grids of photoionised plasmas for the new \textit{NICER+XMM-Newton} data as previously done for the archival \textit{NICER} data in DS23. We adopted a logarithmic grid of ionisation parameters (log\,$\xi$ [erg/s cm] between 1 and 6 with 0.2 steps) and $v_{\rm LOS}$ ranging between $-0.3c$ and {$+0.1c$}. We tested three values of velocity dispersion ($v_{\sigma}$ = 1000, 10,000 and 20,000\,km/s). In Fig.\,\ref{fig:xmm_nicer_pion_scan}, we show the \texttt{pion} scan obtained adopting a velocity dispersion of 20,000\,km/s, which provided the largest improvement corresponding to a remarkable $\Delta\,C=72$ (for 4 additional DOF) with respect to the full ISM model. Such a value would correspond to a detection level well above $5\,\sigma$, even if we take into account the look-elsewhere effect (see, e.g., \citealt{Pinto2023a} and references therein). Given such a large statistical improvement, we refrained from running specific simulations as it would be redundant. The best-fitting solution is found for $\log\,\xi=2.0$ for $v_{\rm LOS}=-0.06c$. In addition, weaker structures can also be seen at $\log\,\xi\gtrsim4$ with $v_{\rm LOS}$ below $\sim-0.1c$, indicating the potential presence of a multiphase plasma.

A direct fit of the data with the \texttt{pion} component in addition to the full ISM model (dubbed the complete or best-fit model) yielded a mildly-relativistic Doppler blueshift $v_{\rm LOS}= -0.061 \pm 0.005 c$ and a consistent broadening $v_{\rm \sigma}= 0.065 \pm 0.005 c$. The other two free parameters gave $N_{\rm H} = (3.0 \pm 0.5) \times10^{21}$\,cm$^{-2}$ and $\log \xi = 1.88 \pm 0.05$ erg s$^{-1}$ cm. 
The spectra, best-fit model, and residuals can be found in Fig.\,\ref{fig:best_fit} (top two panels). Some further very broad continuum-like residuals are still present, most likely due to \textit{NICER}-\textit{XMM-Newton}/EPIC cross-calibration issues between 3-5 keV. 
The results obtained here are in line with those obtained for the intermediate-soft states of this source in DS23. The availability of the high-resolution RGS spectrum enabled us to account for a more complete ISM model (contributing to many rest-frame narrow lines) which resulted in a smaller column density of the \texttt{pion} model with respect to DS23.

\subsection{The archival \textit{NICER} spectra} \label{sec:nicer}

\begin{figure}
   \centering
   \includegraphics[scale=0.725]{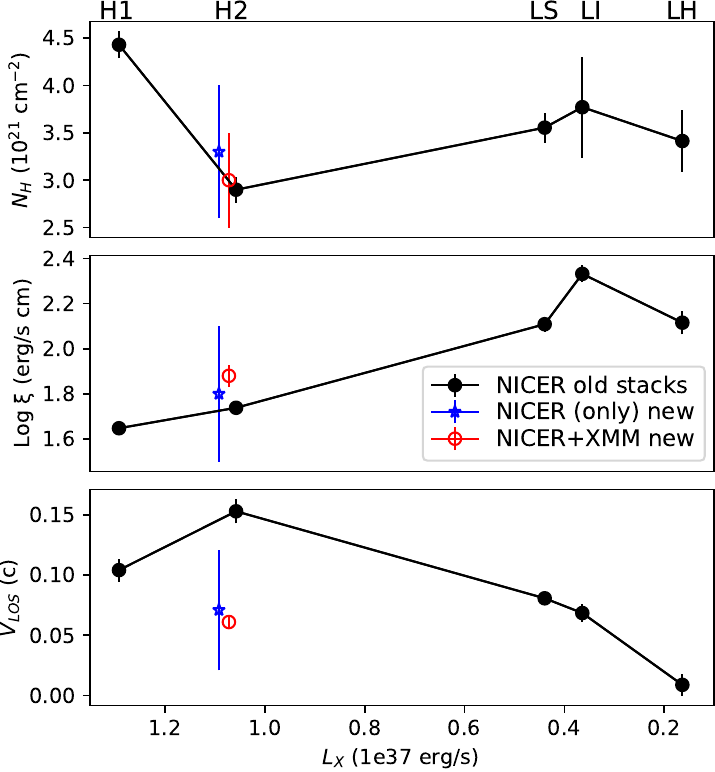}
      \caption{Best-fit parameters of the outflowing plasma component (\texttt{pion}) for the five \textit{NICER} {stacked} spectra sorted according to the intrinsic (unabsorbed) X-ray 0.3-10 keV luminosity (re-fit of the flux-resolved spectra with the RGS ISM model) {as compared with the new RGS (open red circles) and NICER-only (blue star) results}. The X-axis is inverted following the HID evolution from high-soft to low-hard states.}
         \label{fig:NICER_vs_RGS}
   \end{figure}

In DS23, it was shown that the outflow properties were connected to the source spectral state, although the results were obtained by adopting the simple baseline model like the one used in Sect.\,\ref{sec:baseline} due to the limited \textit{NICER} spectral resolution.
A meaningful comparison between the results obtained with the archival \textit{NICER} data and the more recent  \textit{NICER} and, especially, \textit{XMM-Newton} observations, requires the adoption of an identical spectral model. For this reason, we performed a new fit of the \textit{NICER} data presented in DS23 which consisted of five flux- and hardness-resolved spectra. The adopted model is consistent with the best-fit model obtained for the new simultaneous \textit{NICER} and \textit{XMM-Newton} spectra described in Sect.\,\ref{sec:bestfit}. In order to avoid degeneracy, we fixed all the parameters of the full ISM model (which are not expected to vary) and the velocity dispersion of the photoionised absorber to the RGS fit results. In addition to the continuum parameter, the only additional free parameters were the column density, the ionisation parameter, and the line-of-sight velocity of the photoionised absorber. In Fig.\,\ref{fig:NICER_vs_RGS}, we compare the results obtained for the \textit{NICER} archival data (black, filled circles) with those obtained with simultaneous fits of the new data (red, open circles). {A final quick test was performed by considering only the new \textit{NICER} spectrum for comparison with the archival data (see blue star in Fig.\,\ref{fig:NICER_vs_RGS}).} The panels refer to the main parameters of the photoionised absorber; the X-axis shows the X-ray luminosity, sorted from softest to hardest spectral states, as described in DS23. H1 and H2 are two high states with different hardness, while LS-LI-LH refer to low states with soft, intermediate and hard spectra. The results are discussed in Sect.\,\ref{sec:discussion}.

\subsection{The archival \textit{NuSTAR} spectrum} \label{sec:nustar}

\begin{figure}
   \centering
   \includegraphics[scale=0.45]{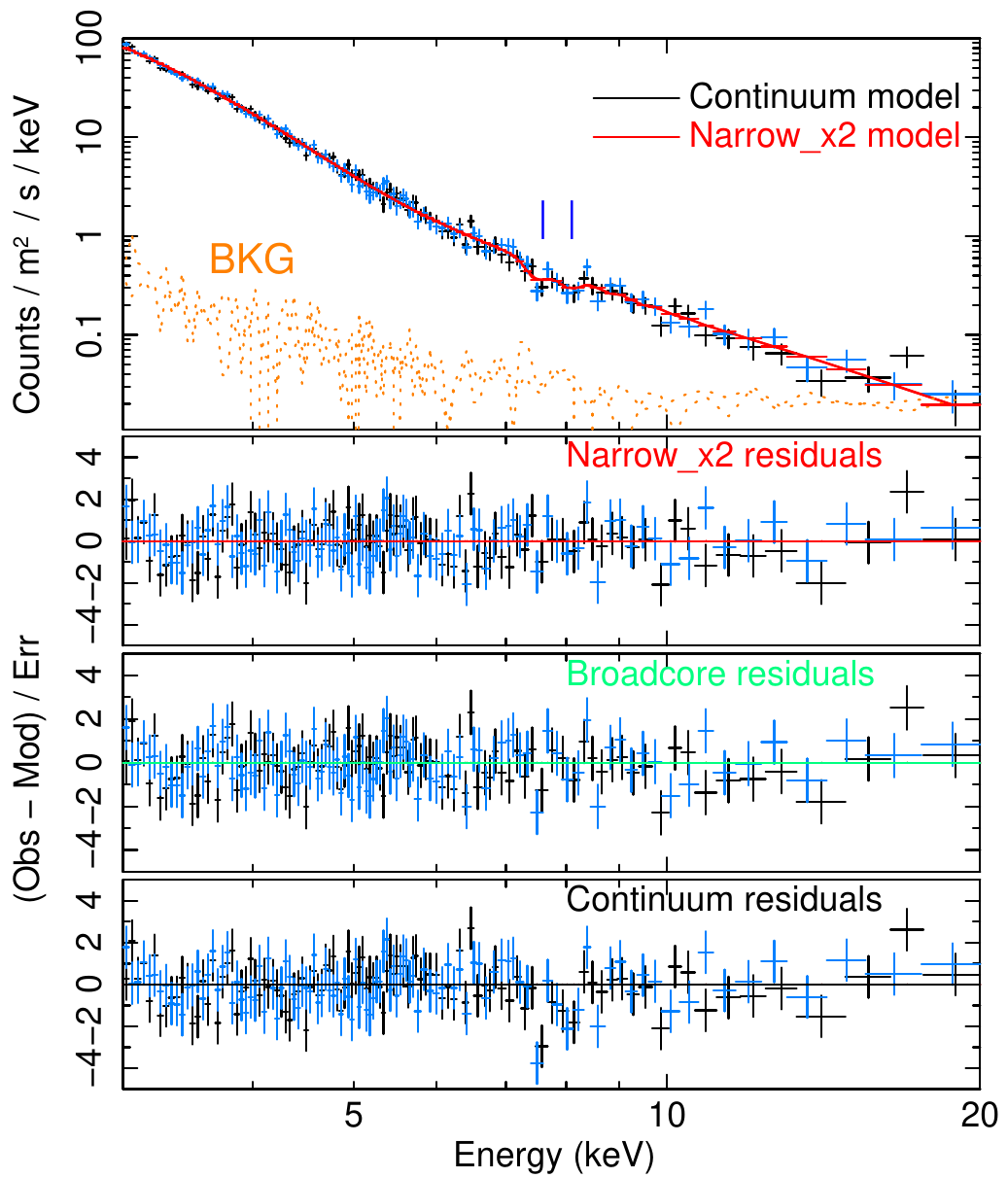}
   \includegraphics[scale=0.53]{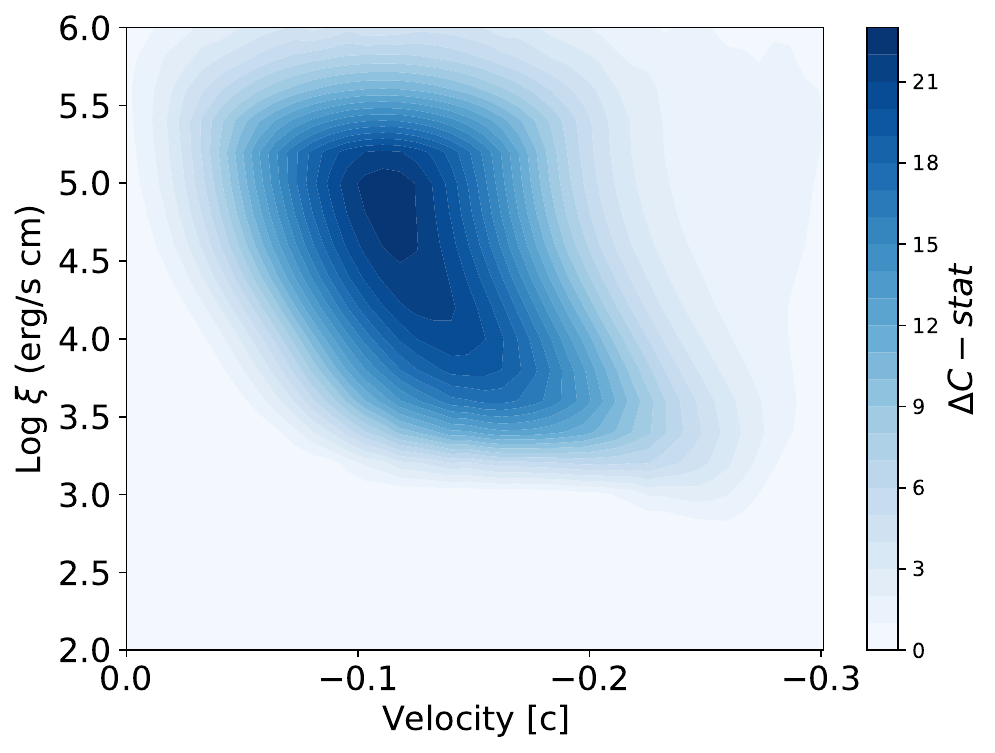}
    \vspace{-0.2cm}
      \caption{{Top panel: \textit{NuSTAR} spectrum (the only one available) for the soft-intermediate state with alternative models.} Bottom panel: model scan for photoionised plasma in absorption applied onto the \textit{NuSTAR} spectrum for a velocity dispersion of 20,000\,km/s.}
         \label{fig:nustar}
   \end{figure}

Unfortunately, at present there is only one observation taken with this telescope and corresponds to a high-soft state. 
This soft state spectrum was studied by {R22} who focussed their study on the spectral continuum.
We noticed some residuals around 8 keV which pushed us to further investigate the same spectra in order to  understand if that feature could be related to the presence of any outflows. The \textit{NuSTAR} FPM\,A/B spectra are shown in Fig.\,\ref{fig:nustar} (upper panel). At first, we fit data from 3 to 20 keV, where the source is above the background and, as above, grouping the spectra with optimal binning in {\scriptsize{SPEX}}. The adopted continuum model was the baseline model (disc-blackbody + Comptonisation) introduced in Sect.\,\ref{sec:baseline}. The residuals to the baseline model are shown in the fourth panel (from top to bottom). Absorption features can be seen between 7.5 and 8.5 keV. The multiphase ISM produces narrow lines predominantly below 3 keV, which allows us to adopt a simple model with only neutral gas for the fit of the \textit{NuSTAR} spectrum.

Motivated by the results obtained with the archival \textit{NICER} data and the RGS confirmation of the detection, we performed a series of model grids for the \textit{NuSTAR} spectra. We used the \texttt{pion} code in {\scriptsize{SPEX}}, which instantaneously computes the photoionisation balance, using the current continuum model as radiation field. As in Sect.\,\ref{sec:bestfit}, we adopted a logarithmic grid of ionisation parameters (log\,$\xi$ [erg/s cm] between 2 and 6 with 0.2 steps) and $v_{\rm LOS}$ ranging between -0.3c and zero. We tested three values of velocity dispersion ($v_{\sigma}$ = 1000, 10,000 and 20,000\,km/s). In Fig.\,\ref{fig:nustar} (bottom panel) we show the \texttt{pion} scan obtained adopting a velocity dispersion of 20,000\,km/s in agreement with the \textit{XMM-Newton}/RGS results. A broad core with $v_{\sigma} =$ 20,000\,km/s provides the largest improvement with $\Delta\,C=23$, which corresponds to about $3.0\,\sigma$, if we take into account the look-elsewhere effect (see, e.g., \citealt{Pinto2023a} and references therein). The best fit is achieved for $v_{\sigma}=0.05 \pm 0.02\,c$, $v_{\rm LOS}=-0.11\pm0.02\,c$ and log\,$\xi=5.0\pm0.5$ and $N_{\rm H} = (1.0 \pm 0.5) \times10^{24}$\,cm$^{-2}$.

In Fig.\,\ref{fig:nustar} (second and third panels from top to bottom), we also show the \textit{NuSTAR} FPM\,A/B spectral residuals computed for two alternative models using a single plasma phase with large velocity dispersion (as obtained from the \texttt{pion} scan) and a two-phase absorber consisting of two \texttt{pion} components with lower broadening ($\lesssim$\,1,000\,km/s) and all parameters coupled with the exception of the line-of-sight velocities ($v_{\rm LOS\,1,2}=-0.1c$ and $-0.15c$, $\log \xi= 3.6$ for a total $N_{\rm H} \sim 2.5 \times 10^{22}$\,cm$^{-2}$). The use of two narrow-line \texttt{pion} components provides a comparable improvement ($\Delta\,C=26$) with respect to the baseline continuum model, most likely due to the limited spectral resolution.

\vspace{-0.2cm}

\section{Discussion} \label{sec:discussion}

The discussion is structured as follows. At first, we compare the ISM properties with those in the literature to corroborate our results and obtain some further indication of the source location within the Galaxy. Then we compare the results obtained with different instruments at different epochs. Finally, we provide some more properties of the outflowing plasma and an attempt to understand its nature.

\subsection{The ISM along the line-of-sight}\label{sec:ism}

From the \texttt{hot} component describing the cold, neutral, interstellar gas we can compute the column densities of individual atomic species (for the adopted Solar abundances). These can be compared (and summed) with those of ions and solids as measured with the \texttt{slab} and \texttt{amol} components, respectively. By summing up the contributions from each component, we obtained the following abundances in Solar units: O/H = Ne/H = $1.5\pm0.1$, Mg/H = $1.6\pm0.2$, Si/H = $1.2\pm0.3$, and Fe/H = $1.2\pm0.1$. Super-Solar values are expected for a line-of-sight that crosses the Galactic inner regions due to the metallicity gradient, but our estimates exceed by about 20\% those found in sources at distances below 8\,kpc from the Sun \citep{Pinto2013,Zeegers2019}. This would confirm a larger distance for J1810 that was previously suggested by {R22}, using the Galactic mass density model of low-mass XRBs \citep{Grimm2002} as
implemented in \cite{Atri2019}, and by DS23 (about 20-30 kpc), based on the estimation of the accretion rate of $0.1 \, \dot{\rm M}_{\rm Edd}$.

Comparing the column densities of each element or ion as obtained from the various components, we estimate that the oxygen is roughly distributed as follows: $10-15\,\%$ in solids, $75-80\,\%$ in the neutral phase, $10-15\,\%$ in the low-ionisation (O\,{\sc ii-iv}) phase, and $5-10\,\%$ in the high-ionisation (O\,{\sc vii-viii}) phase. The ratios between the gaseous phases are consistent for oxygen and neon (the latter is not depleted into solids). For heavier elements, we find a larger depletion with solids contributing to approximately $40-50\,\%$ of neutral iron and, at least, $80\,\%$ and $90\,\%$ of magnesium and silicon, respectively. Our results on ionic fractions and dust depletion are fully consistent with recent results on several samples of Galactic XRBs (see, e.g., \citealt{Pinto2013,Zeegers2019,Rogantini2020,Psaradaki2024}). This also implies that we do not find evidence of further narrow absorption lines from slow-moving plasma in the circumstellar medium around J1810.

\subsection{Comparison between different epochs}\label{sec:comparisons}

J1810 was discovered as a soft X-ray transient in 2018 and since then has been randomly {moving from soft (disc-dominated) to hard (corona-dominated) states} by passing through intermediate states where both the disc and the corona contributed significantly to the observed bolometric flux (see Fig.\,\ref{fig:swift}, {R22} and Fig. 2 in DS23). 

DS23 discovered evidence for a strong 1\,keV absorption feature that was more prominent in the soft states. This, when described with a photoionisation model, indicated fast outflows, achieve $0.1\,c$ in the soft states. Our simultaneous observations successfully triggered with \xmm\ and \nicer\ caught J1810 in a high-soft state with a typical disc temperature of about 1 keV and less than $10\,\%$ of the $0.3-10$\,keV flux provided by the Comptonisation, in which we would have expected a strong 1\,keV line. This was indeed confirmed in all instruments, including \xmm\ EPIC (despite some calibration issues for the latter, see Fig.\,\ref{fig:continuum_fit} and \ref{fig:epic_calib}). A deep exploration of the ($\xi,v_{\rm LOS}$) parameter space was performed with photoionisation models and confirmed previous results obtained with archival data, but also places a strong constraint on the line broadening or velocity dispersion and LOS velocities, both of about $6\,\%$ of the speed of light. This was not within the reach of the archival data as only X-ray gratings are currently able to resolve spectral features in the soft X-ray band ($0.3-2$\,keV, see Fig.\,\ref{fig:xmm_nicer_pion_scan}). This was also corroborated by the application of a comprehensive and up-to-date ISM absorption model (see Fig.\,\ref{fig:best_fit}). We emphasise that if the 1\,keV line was composed of several absorption features, on the one hand, they would be distinguished with the RGS, on the other hand, they would have been easily fitted through the \texttt{slab} model.

For a comparison with different epochs, we performed a new modelling of the archival \nicer\ flux/hardness resolved spectra that were first published in DS23. The results obtained for the new \nicer\ + \xmm\ data generally agree with those shown by the same high/intermediate-soft spectral state with the exception of the outflow velocity, which was lower by a factor of 2-3 in the new data (see Fig.\,\ref{fig:NICER_vs_RGS}). This indicates some evolution of the outflow (the H2 archival spectrum was obtained by averaging \nicer\ spectra with similar flux and hardness levels). {This is confirmed by the broad agreement between the results obtained with the individual fits of the new NICER and RGS spectra.}

Motivated by such a strong detection, we reanalysed the \textit{NuSTAR} spectrum presented in {R22} and noticed a possible absorption feature near 8\,keV (see Fig.\,\ref{fig:nustar} top panel), which could even have been  spotted although not mentioned in that paper. The application of the same photoionisation model and the exploration of the parameter space ($\xi,v_{\rm LOS}$) enabled the detection at about $3.0\,\sigma$ of an outflowing plasma with a slightly larger speed ($0.1\,c$, see Fig.\,\ref{fig:nustar} bottom panel). This would correspond to a canonical UFO as those seen in AGN at different Eddington ratios whose nature is still highly debated (see, e.g., \citealt{Tombesi2010,Gianolli2024}). However, the spectral resolution of \textit{NuSTAR} is not high enough to distinguish between a single broad-core line and multiple narrow lines. The presence of absorption in the Fe\,K band is confirmed by the EPIC data whose photoionisation grids show a secondary hotter solution with log\,$\xi\sim4.5$ in Fig.\,\ref{fig:xmm_nicer_pion_scan}. The ($\xi,v_{\rm LOS}$) map would favour a slow-moving plasma, perhaps even consistent with being at rest but given the relevance of the background above 7\,keV in the timing-mode EPIC spectra (see Fig.\,\ref{fig:best_fit}) we refrain from making further speculations although the comparison between the \textit{NuSTAR} (2018) and the \xmm\ (2023) data would indeed suggest an evolution over time. 
{For instance, the \textit{NuSTAR} observation is characterised by a slightly lower flux (about 50\% less in the 3-10\,keV shared with \xmm\ and \textit{NICER}) and softer SED (the Comptonisation contributes less than 10\% to the 3-10\,keV flux, i.e. three time less than during the \xmm\ observation), which along with the better data quality around 8 keV, might have had an impact on the line detectability.}
Finally, the unavailability of high-resolution data covering the whole X-ray band from 0.3 to 10 keV does not allow us to distinguish between a continuous absorption measure distribution (or a stratified plasma) and a more discrete multiphase outflow. A simultaneous \xmm\ RGS + \textit{XRISM} Resolve observation would be needed for such a scope.

\subsection{The nature of the outflow} \label{sec:outflow_nature}

Following previous calculations for the flux/hardness-resolved spectra performed in DS23, we computed some properties of the outflow from the best-fit solution describing soft X-ray absorption lines ($v_{\rm LOS}= -0.061\,c$,\, $N_{\rm H} = 3.0  \times 10^{21}$\,cm$^{-2}$ and $\log \xi = 1.88$ erg s$^{-1}$ cm). From the definition of ionisation parameter we can compute a maximum distance for the photoionised plasma: $r_{\rm max} \sim L_{\, \rm ion}/(\xi\,N_{\rm H}) \sim 10^7 \,R_{G}$ where for the gravitational radius, $R_{G}$, we adopted a typical $M_{\rm BH} = 10$\,\Msun. A lower limit on the distance can be obtained from the escape radius $r_{\rm min} \sim 2 G M_{\rm BH}/v^2 = 500\,R_{G}$. These radii may be used to estimate a range of plasma densities: $n_{\rm \, H, \,min} \sim N_{\rm H}/r_{\rm max} = 5 \times 10^7$\,cm$^{-3}$ and $n_{\, \rm H, \, max} \sim N_{\rm H}/(b\,r_{\rm min}) = 4 \times 10^{13}$\,cm$^{-3}$ (for an adopted volume filling factor $b=0.1$). Such a range of densities, although very broad, is comparable with those found with direct (He-like triplets) or indirect (variability) arguments in outflows observed in ULXs (e.g., \citealt{Pinto2023a}), AGN (e.g., \citealt{Xu2024}) and Galactic XRBs (e.g., \citealt{Psaradaki2018, Kosec2024}). {This estimate is conservatively based on thermal driving, an MHD wind could result in an even smaller radius and larger density.}

The outflow rate can be expressed as $\dot{M} = 4 \, \pi \, R^2 \, \rho \, v \, \Omega \, C$ where $\Omega$ and $C$ are the solid angle and the volume filling factor (or \textit{clumpiness}), respectively, and $R$ is the plasma distance from the source. The kinetic power of the winds is $L = 0.5 \, \dot{M} \, v^2 = 2 \, \pi \, m_p \, \mu \, \Omega \, C \, L_{\rm ion} \, v^3 \, / \, \xi \sim 0.06$ where we used $\xi=L_{\rm ion}/n_{\rm H}R^2$. The largest uncertainties concern the solid angle and the clumpiness.
In agreement with former results in DS23, the outflow rate appears mildly super-Eddington ($\dot{M} \sim 2 \, \dot{\rm M}_{\rm Edd}$, assuming $C=0.1$, otherwise Eddington-limited for $C\lesssim0.05$). In DS23 it was reported that the outflow could remove most of the material, perhaps explaining the transition to low states where both the accretion and outflow rates are significantly smaller (0.1-0.2 $\dot{\rm M}_{\rm Edd}$). Moreover, such a value would still require an accretion rate of $0.1 \, \dot{\rm M}_{\rm Edd}$, which would correspond to an intrinsic luminosity of $10^{38}$ erg/s for a BH, indicating that the source is even more distant at $\gtrsim20$\,kpc, as also suggested by DS23 and {R22}. For the kinetic power, we obtain $ L \sim 0.06 \, L_{\rm Edd}$ which together with the large outflow rate could indicate that the mass load is high, thus slowing down the flow. 

The application of the full ISM model (in substitution of a simple neutral gas) confirmed the trend for the parameters of the photoionised plasma as previously obtained in DS23. In particular, while  the velocity decreases, the ionisation parameter increases towards the hard state, which is associated to X-ray ($0.3-10$\,keV) and bolometric ($10^{-3}-10^{\,3}$\,keV) luminosities lower by factors of about 5 and 2, respectively, with respect to the hard state. Although this {might} appear counter-intuitive for $\xi$, we notice that the strong hardening of the SED may result into a substantial increase of the heating  and, therefore, of $\xi$.

The decrease in the velocity might be investigated in the context of magnetically driven winds. For example, a local heat deposition in the  upper layers of the disc gives rise to an enhanced mass loss rate (see, e.g., \citealt{Casse2000b}). This extra thermal effect leads to an increase of the load put on the field lines and therefore a decrease of the wind velocity (which would explain the trend towards the harder states). Alternatively, as the ionising flux increases, the innermost streamlines get ionised first, making us probing more distant outflowing regions, although a substantial decrease in the column density should be foreseen.

The mildly relativistic velocities as measured for the Doppler shift and broadening in the soft and intermediate states is already a strong indication for magnetic drive because in an Eddington-limited regime the thermal mechanism would be very inefficient in driving outflows significantly above 1000\,km/s as shown by global hydrodynamic simulations (see, e.g., \citealt{Done2018}). The anticorrelation between $v_{\rm LOS}$ and $\xi$ is also not expected in thermal winds, while it could be reconciled with MHD scenario of certain density profile {(see below for more detail and \citealt{Fukumura2017})}. In the latter case, for instance, the extra heating on the disc surface (observed during the hard state) would enhance mass loading onto magnetised winds, making it heavier and thus slower. Magnetic winds also predict rather broad opening angles with cooler components coming from outer regions which would then produce visible absorption lines in the case of spectral hardening (see previous comment on the probing of more distant outflowing regions and Fig.\,\ref{fig:NICER_vs_RGS}). We notice that such an anticorrelation between $v_{\rm LOS}$ and $\xi$ is also observed in UFOs detected in AGN (see, e.g., \citealt{Xu2021,Gianolli2024}). The description of lower velocities in terms of outer launching regions has also been invoked  to explain the appearance of slower outflow components in ultra-luminous X-ray sources (see, e.g., \citealt{Pinto2020b}).

   \begin{figure*}
   \centering
   \includegraphics[scale=0.3]{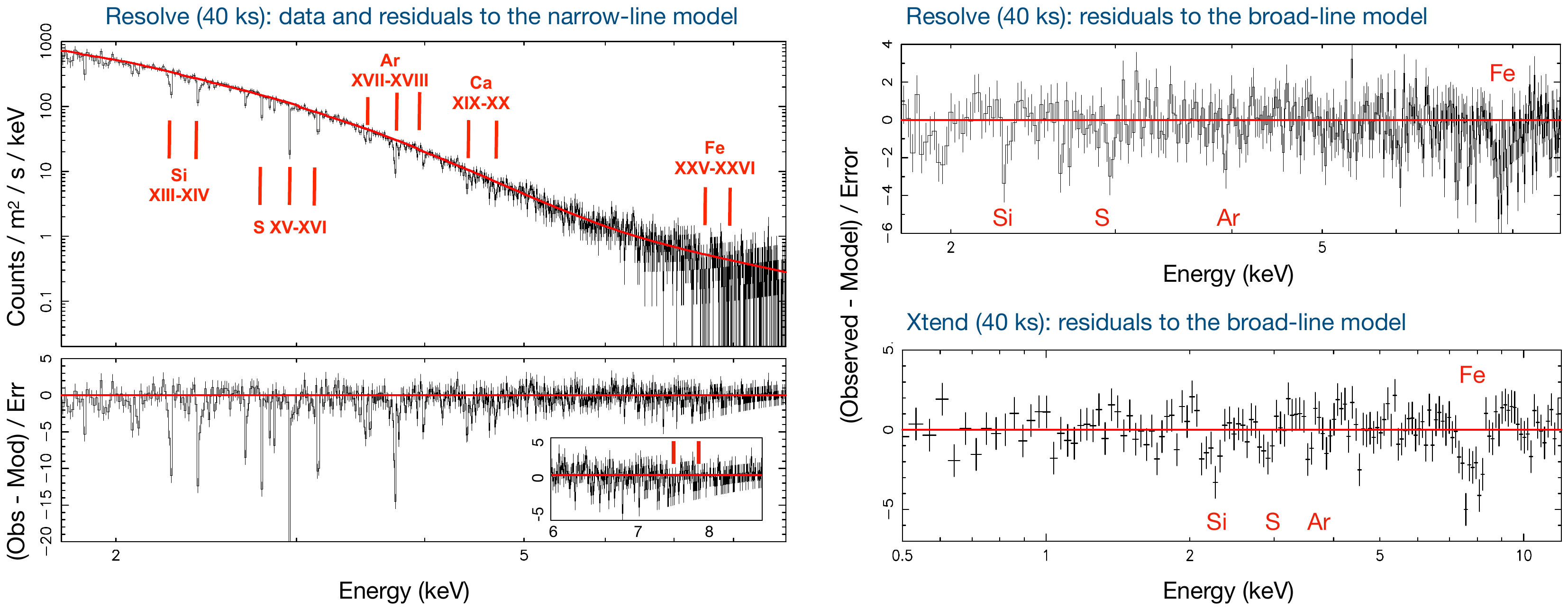}
      \caption{Left panel: {\textit{XRISM}} Resolve simulation (40\,ks) of J1810 with a photoionised absorption model (two narrow components). {Resolve gate-valve closed was adopted.} Top-right panel: Resolve simulation with a broad-line model. Bottom-right panel: {\textit{XRISM}} Xtend simulation with the broad line model simulation. In both panels the \texttt{pion} component has been removed to show that the strongest lines are still detected.}
         \label{fig:xrism}
   \end{figure*}

{The velocity dispersion, as estimated with a single photoionised component, for the cool absorber is rather extreme even for a magnetic field. This is indeed difficult to achieve as velocity shear unless the flow is very extended (i.e. no clumpiness at odds with recent XRISM results on different sources involving less than a few 1000 km/s). It is plausible that the current value of 20,000 km/s could be masquerading a broad feature behind a sub-structure of multiple, narrow features just like PDS 456 and PG 1211+143 (\citealt{XrismCollaboration2025,Mizumoto2026}). This would falsely imply that the outflow is stratified and extended. XRISM / Resolve resolution at 8 keV (Fe K 0.1c outflows) exceeds E/dE~1600, which is a factor 8x better than the high-counts RGS 1st order at 1 keV. It is likely that the 1 keV broad feature is a merge of multiple lines with a lower broadening around 1000-2000 km/s which we cannot fully resolve. Equally good fits can be achieved with a series of narrow PION components (at different LOS velocities, see 2D maps in Fig. 4) just as it occurs for the Fe K region observed with CCDs.}

For the column density $N_{\rm H}$, to be weakly sensitive with distance, the outflow would need to have a narrow cone and the source to be seen at sufficiently low inclinations. This is expected by the non-detection of slow-moving Fe\,K thermal winds in the \nustar\ spectrum, the lack of QPOs and the strong radio RMS versus X-ray RMS (a factor 10:2, likely due to variable Doppler boosting seen at a low inclination, see {R22}). {More in detail, assuming a continuous density distribution that scales as $n \sim r^{-p}$ (and ignoring a clumpy nature for simplicity), would result in $N_{\rm H} \sim r^{(1-p)}$. Therefore, a shallow density profile around $p \sim 1$ would produce $N_{\rm H}$ that is only weakly sensitive to radius, allowing a nearly constant column; this would broadly agree with Fig.\,\ref{fig:NICER_vs_RGS}. Such $p$ values correspond to a large disc ejection efficiency (see, e.g., \citealt{Chakravorty2016,Datta2024}), which is expected from weakly magnetised discs. Therefore, our results would be consistent with the fact that the outer regions would be weakly magnetised, while the innermost ones would be magnetically saturated, namely in the Jet Emitting Disc state. This fits within the JED-SAD framework (\citealt{Ferreira2006, Petrucci2010,Marcel2022}), where it is assumed that the accretion disc sets in a hybrid magnetic configuration.}

\subsection{Prospects}\label{sec:prospects}

As previously pointed in Sect.\,\ref{sec:comparisons}, we currently lack high-spectral-resolution coverage of the whole $0.3-10$\,keV energy band where most ionic absorption lines from X-ray outflows are expected, which prevents us from obtaining a complete view of the outflow properties. To showcase the capabilities of the {\textit{XRISM}} observatory {(with closed Resolve gate-valve)}, we have performed a set of simulations for exposure times ranging from 10 to 100\,ks by adopting either the single broad core or the two narrow \texttt{pion} components that provide comparable descriptions of the Fe\,K blueshifted feature observed in the \nustar\ spectrum (see Sect.\,\ref{sec:nustar}). In Fig.\,\ref{fig:xrism} (left panel), we show a {\textit{XRISM}} / Resolve simulation (at 5\,eV resolution) with the two narrow-line model. The \texttt{pion} absorption lines were removed from the model to highlight the expected residuals, particularly in the intermediate band (2--5 keV). This looks very similar to those produced by the strong outflow observed in GRO\,J1655--40, which was also interpreted as a magnetic wind (see, e.g., \citealt{Miller2006Nature}). 

In Fig.\,\ref{fig:xrism} top-right (bottom-right) panel shows the residuals to the Resolve (Xtend) spectrum simulated with the broad line model where \texttt{pion} has been removed. For our simulations we adopted a conservative low-soft state with $f_{\rm 2-10 \, keV} = 1 \times 10^{-10} \rm erg \, s^{-1} cm^{-2}$, which represents the lower-flux end of the soft-intermediate states (see also Fig.\,\ref{fig:swift} and \ref{fig:NICER_vs_RGS}). In the brighter soft / intermediate states (as for the \xmm\ data studied in this work) we expect an even higher flux (up to a factor 3) in the Fe\,K region. According to our photoionisation calculation, the Fe\,{\scriptsize XXV-XXVI} absorption lines (along with some of the sulphur, calcium and argon lines at lower energies) will remain comparably strong in the intermediate state, while in the hard state the X-ray heating is high enough to fully ionise most ions.  

According to our simulations, in a Resolve observation of 40\,ks, the {\texttt{pion}} component produces a spectral improvement $\Delta C = 60$ for the pessimistic case (broad core), corresponding to a confidence level $>5.0\,\sigma$ \citep{Pinto2021} and much higher at a lower velocity dispersion. We expect an accuracy of better than 20\% for the {\texttt{pion}} column density and ionisation parameter and much smaller for the velocities. This is sufficient to distinguish between the two solutions and measure the outflow kinetic rate. In the pessimistic case, the precision on the S/Fe (Si/Fe or Ar/Fe) abundance ratio will be $\sim30$\,\% ($\sim60$\,\%). The Xtend data will further corroborate the detection and, importantly, constrain the shape of the broadband spectrum, which is crucial for photoionisation balance calculation.

\section{Conclusions}

In this work, we performed a high-spectra-resolution study of the notable outflow observed in MAXI\,J1810-222. The \textit{XMM-Newton} / RGS spectrum confirmed (with a higher accuracy) the presence of a mildly relativistic wind that was previously suggested by \textit{NICER} {and would favour magnetically-driven winds although thermal effects may still contribute to mass loading}. Measurements of Doppler shifts and velocity dispersion indicate that the lines are intrinsically broad. \textit{NuSTAR} and \textit{XMM-Newton} / EPIC spectra also show a further hotter component that produces additional absorption in the Fe\,K band ($7.5-8.5$\,keV) suggesting a stratified or a multiphase outflow. \textit{XRISM} observations of J8110 will complement the \textit{XMM-Newton} / RGS data through an unprecedented view of the Fe\,K band that will enable the first comprehensive, high-resolution study of the complex extreme outflow in this source and shed new light on outflow mechanisms in XRBs.

\section*{Data availability}

All data and software used in this work are publicly available from the ESA and NASA Science Archives (\url{https://www.cosmos.esa.int/web/xmm-newton/xsa}), (\url{https://heasarc.gsfc.nasa.gov/}). Our codes are publicly available on GitHub (\url{https://github.com/ciropinto1982}).

\begin{acknowledgements}
CP, MDS, AD and FP acknowledge funding from European Union - Next Generation EU, Mission 4 Component 1 CUP C53D23001330006 (PRIN MUR 2022 SEAWIND project No. 2022Y2T94C) and INAF Large Grant 2023 BLOSSOM O.F. 1.05.23.01.13. {MDS and AD acknowledge ASI-INAF program I/004/11/6 (Swift). POP, JF acknowledge financial support from the french spatial agency CNES, and from the Action Thématique "Phénomènes Extrêmes et Multimessagers" from the Astronomy \& Astrophysics french National Program of CNRS. MP acknowledges support from the JSPS PF for Research in Japan, grant number P24712, and JSPS Grants-in-Aid for Scientific Research-KAKENHI, grant number J24KF0244. T.M.-D. acknowledges support by the Spanish \textit{Agencia estatal de investigaci\'on} via PID2021-124879NB-I00 and PID2024-161863NB-I00. AM is supported by ERC Consolidator Grant “MAGNESIA” No. 817661 and the National Spanish grant PID2023-153099NA-I00.}
\end{acknowledgements}
   
\bibliographystyle{aa}
\bibliography{ref.bib}

\appendix

\section{{\xmm\ EPIC calibration issues}} \label{sec:epic_calib}

\begin{figure}
   \centering
   \includegraphics[scale=0.3]{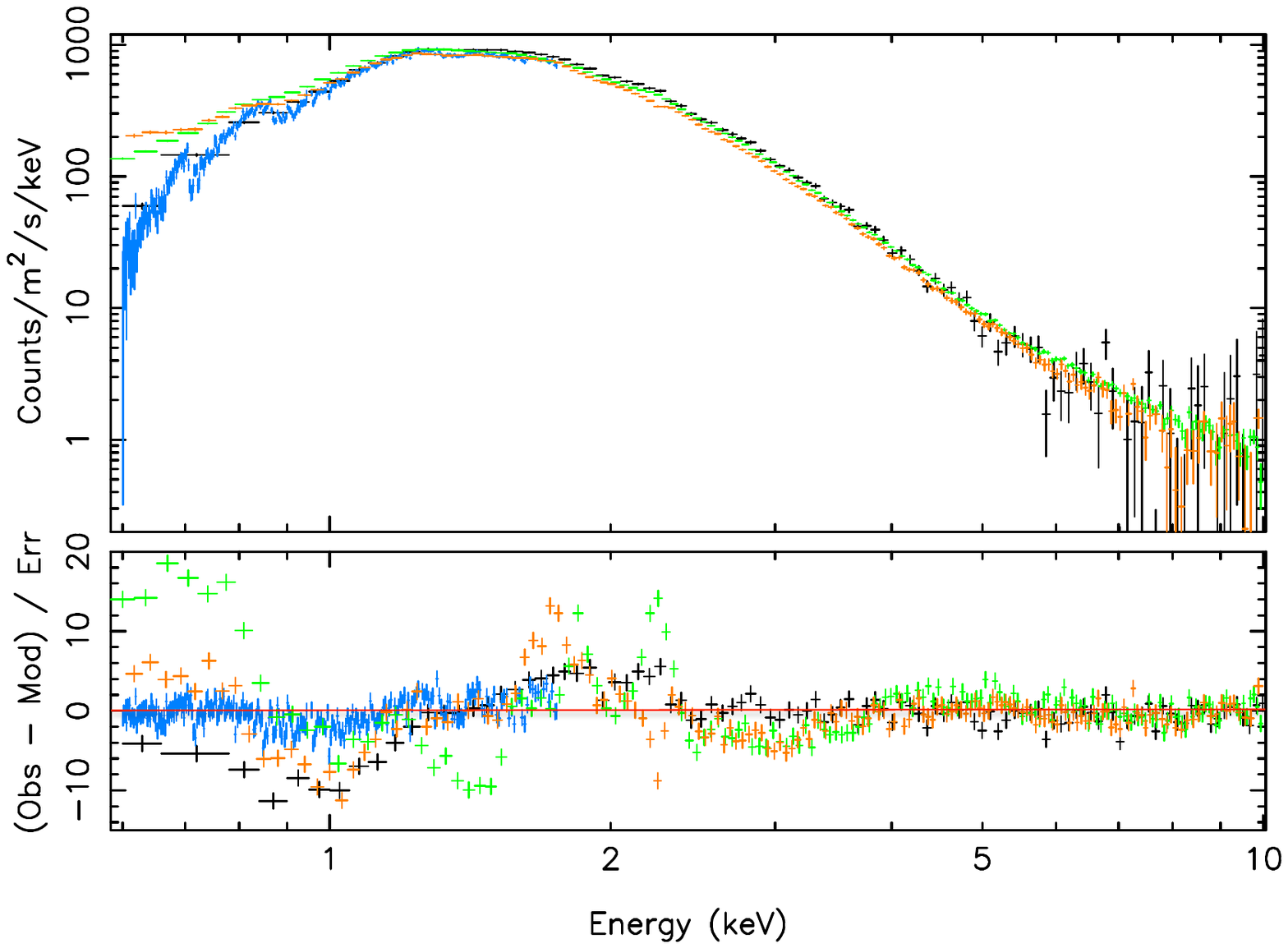}
      \caption{{XMM / RGS (blue data) + EPIC (pn/green and MOS2/orange data) + \textit{NICER} (black data) spectra and  residuals to a simple continuum model. Note EPIC MOS\,2 and pn inconsistencies below 2.6 keV.}}
         \label{fig:epic_calib}
   \end{figure}

{In Fig.\,\ref{fig:epic_calib}, we show a simultaneous fit of the \xmm\ EPIC MOS\,2+RGS and \nicer\ spectra. In this preliminary fit we used a simple continuum model consisting of disc-blackbody and Comptonisation, both absorbed by the ISM neutral gas (see Sect.\,\ref{sec:baseline}). No gaussian lines were included because we wanted to show all instrumental effects. The models are identical for the four spectra with exception of the multiplicative constant, which was fixed to one for \nicer\ and is free for the other spectra.} 

{The \nicer\ spectrum shows the usual edge near 2.2\,keV due to mirror calibration. The EPIC MOS\,2 and pn spectra show inconsistencies between 1.5 and 2.5 keV due to instrumental lines and calibration issues. The \nicer\ and RGS spectra are consistent. Nonetheless, all instruments highlight absorption around 1\,keV. Interestingly, the parameters of the spectral model are consistent within the uncertainties with the baseline model where EPIC data was ignored below 2.6\,keV with the exception for the total X-ray luminosity, $L_{\, \rm 0.5-10\,keV} = (8.6 \pm 0.1)\times10^{36}$\,erg\,s$^{-1}$, and the ISM column density, $N_{\rm H , \, ISM}=(5.9 \pm 0.1)\times10^{21}$\,cm$^{-2}$, both slightly smaller.}

\section{{Search for pulsations}} \label{sec:pulsations_search}

{We searched the \xmm\ EPIC-pn data for periodic signals following the procedure described in \citet{Rodriguez2020}, using the Pulsation Accelerated Search for Timing Analysis (PASTA) software. We corrected the times of arrival (ToAs) of the source events over a grid of $\sim$10,000 trial values, applying a factor of $-\tfrac{1}{2}(\dot{P}/P),t^{2}$. The  range explored was $7\times10^{-6} < |\dot{P}/P;(\mathrm{s}^{-1})| < 1\times10^{-11}$, where $\dot{P}$ is the first derivative of the pulsation period (see \citealt{Rodriguez2020} for further details).
No significant coherent signals were detected. We derived a 3$\sigma$ upper limit of $<0.9\%$ on the pulsed fraction (PF), where PF is defined as the semi-amplitude of a sinusoidal modulation divided by the average count rate.}

\end{document}